\documentclass[iop]{emulateapj}
\usepackage{amsmath}

\newcommand{\rkl}[1]{\left(#1\right)}

\newcommand{\skl}[1]{\left\langle#1\right\rangle}

\newcommand{\dif}{\mathrm{d}}

\newcommand{\pdf}{\ensuremath{\mathrm{P}}}
\newcommand{\pdfv}{\ensuremath{\mathrm{P}_V}}
\newcommand{\pdfm}{\ensuremath{\mathrm{P}_M}}
\newcommand{\gpcc}{\ensuremath{\mathrm{g\,cm}^{-3}}}

\usepackage{color}

\begin{document}

\title{On the evolution of the density pdf in strongly self-gravitating systems}

\author{Philipp Girichidis\altaffilmark{1,2,3,4}}
\author{Lukas Konstandin\altaffilmark{2}}
\author{Anthony P. Whitworth\altaffilmark{4}}
\author{Ralf S. Klessen\altaffilmark{2}}
\affil{$^1$Max Planck Institut f\"{u}r Astrophysik, Karl-Schwarzschild-Str. 1, 85741 Garching, Germany}
\affil{$^2$Heidelberg University, Zentrum f\"{u}r Astronomie, Institut f\"{u}r Theoretische Astrophysik, Albert-Ueberle-Str. 2, 69120 Heidelberg, Germany}
\affil{$^3$Hamburger Sternwarte, Gojenbergsweg 112, 21029 Hamburg, Germany}
\affil{$^4$School of Physics \& Astronomy, Cardiff University, 5 The Parade, Cardiff, CF24 3AA, United Kingdom}
\email{philipp@girichidis.com}

\begin{abstract}
The time evolution of the probability density function (PDF) of the mass density is formulated and solved for systems in free-fall using a simple appoximate function for the collapse of a sphere. We demonstrate that a pressure-free collapse results in a power-law tail on the high-density side of the PDF. The slope quickly asymptotes to the functional form $\pdfv(\rho)\propto\rho^{-1.54}$ for the (volume-weighted) PDF and $\pdfm(\rho)\propto\rho^{-0.54}$ for the corresponding mass-weighted distribution. From the simple approximation of the PDF we derive analytic descriptions for mass accretion, finding that dynamically quiet systems with narrow density PDFs lead to retarded star formation and low star formation rates. Conversely, strong turbulent motions that broaden the PDF accelerate the collapse causing a bursting mode of star formation. Finally, we compare our theoretical work with observations. The measured star formation rates are consistent with our model during the early phases of the collapse. Comparison of observed column density PDFs with those derived from our model suggests that observed star-forming cores are roughly in free-fall.
\end{abstract}

\section{Introduction}%

The density PDF is a powerful tool for analysing astrophysical systems; in both non-gravitating and strongly self-gravitating systems, it reveals key aspects of the underlying physical processes. For supersonic non-gravitating turbulent gas in an isothermal environment the density PDF is log-normal \citep{Vazquez94, PadoanEtAl1997, PassotVazquezSemadeni1998, Klessen2000, KritsukEtAl2007, LemasterStone2008, Federrath08, Federrath10b, PriceFederrath2010, KonstandinEtAl2012}. When self-gravity becomes important, the probability of finding dense regions increases and a power-law tail develops on the high-density side of the PDF \citep{Klessen2000, SlyzEtAl2005, VazquezSemadeniEtAl2008,ChoKim2011,KritsukEtAl2011,CollinsEtAl2012}.\footnote{In the sequel, we shall often refer to the power-law tail that develops on the high-density side of the PDF simply as ``the tail'', and to its logarithmic slope simply as ``the slope''.} Recent observations of column-density PDFs support these theoretical predictions \citep{KainulainenEtAl2009, FroebrichRowles2010, LombardiEtAl2010, SchneiderEtAl2012}. 

The density PDF can be used to evaluate key aspects of star formation, like the efficiency and the stellar initial mass function (IMF) \citep[see, e.g.,][]{Krumholz05,PadoanEtAl1997,HennebelleChabrier2008,ShadmehriElmegreen2011,VeltchevEtAl2011,DonkovEtAl2011,DonkovEtAl2012}.

A possible theoretical relation between the log-normal character of the density PDF in turbulent media and the appearance of a power-law tail at high densities has been developed by \citet{Elmegreen2011}, using convolution PDFs that depend on the ratio of maximum to minimum average cloud density, i.e., the core-to-edge average density ratio.

In this paper we develop a model for the evolution of the density PDF, based on free-fall collapse. We start with the initial density PDF of a gaseous system, not specifying its spatial structure, nor how it evolved to this state. We then evolve the system in time with two approximations. In the first one we apply the free-fall analysis directly to the density PDF. In the second approach we assume that the system can be described by an ensemble of free-falling homogeneous spheres, such that the subsequent time evolution is solely determined by the collapse of the individual spheres.

Using the time evolution of the PDF, we derive the free-fall accretion rate, defined as the rate at which gas evolves above an arbitrary threshold density. This accretion rate can then be tuned to serve as an estimate for the star formation rate.

The model is compared to observations of star-forming regions taken from \citet{KainulainenEtAl2009, HeidermanEtAl2010} and \citet{SchneiderEtAl2012}. The observed star formation rate is consistent with the values of the free-fall model at early stages of the collapse. Comparing the observed column density PDFs with our model suggests that the star-forming cores are in free-fall.

\section{Analytic Collapse Model}%
\label{sec:sec-analytic-collapse}

\begin{figure}
  \centering
  \includegraphics[width=8cm]{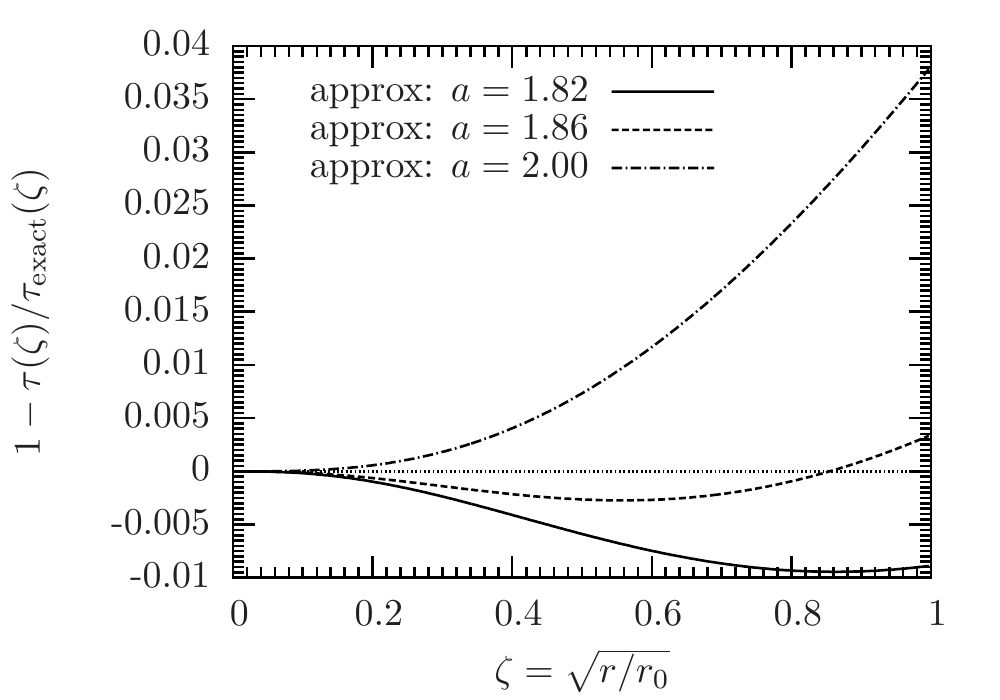}
  \caption{Fractional difference between the approximate solution of equation~\ref{eq:approx-ff-solution} and the exact solution of equation~\ref{eq:exact-ff-solution}.}
  \label{fig:error-approx-solution}
\end{figure}

We begin with a review of the analytic solution for a uniform-density sphere in free-fall. The time evolution of the density can then be described by a simple approximate function and applied to the density PDF. The radius, $r$, of such a sphere obeys
\begin{equation}
  \label{eq:free-fall-DGL}
  \frac{\dif^2r}{\dif t^2} = -\,\frac{GM}{r^2} = -\,\frac{4\pi G\rho_0 r_0^3}{3r^2}\,.
\end{equation}
Here, $G$ is Newton's constant, $M$ is the mass of the sphere, $\rho_0$ and $r_0$ denote the initial density and radius of the sphere at time $t\!=\!0$, and the collapse starts from rest. Using the free-fall time,
\begin{equation}
  t_\mathrm{ff} = \sqrt{\frac{3\pi}{32 G \rho_0}\,},
\end{equation}
as the characteristic time, and defining the dimensionless radius $\zeta=r/r_0$, and the dimensionless time $\tau_\mathrm{exact}=t/t_\mathrm{ff}$, one obtains the equation,
\begin{eqnarray}
  \frac{\dif^2\zeta}{\dif\tau_\mathrm{exact}^2}&=&-\,\frac{\pi^2}{8\zeta^2},
\end{eqnarray}
with initial conditions $\zeta\!=\!1$ and $\dif\zeta/\dif\tau_\mathrm{exact}\!=\!0$ at $\tau_\mathrm{exact}\!=\!0$. This can be solved to give
\begin{equation}
  \label{eq:exact-ff-solution}
  \tau_\mathrm{exact} = \frac{2}{\pi}\rkl{\arccos\sqrt{\zeta} + \sqrt{\zeta(1-\zeta)}}
\end{equation}
\citep[e.g.,][]{Hunter1962,Tohline1982}. This equation can not be inverted analytically, but a good approximation for $\zeta(\tau)$ can be obtained by setting
\begin{equation}
  \label{eq:approx-ff-solution}
  \tau = \sqrt{1-\zeta^{3/a}}.
\end{equation}
Equation~\ref{eq:approx-ff-solution} deviates least from the exact solution (equation~\ref{eq:exact-ff-solution}) if we substitute $a=1.8614$, and figure~\ref{fig:error-approx-solution} shows the error, $1-\tau/\tau_\mathrm{exact}$, that results from this and other substitutions. Inverting equation~\ref{eq:approx-ff-solution} yields
\begin{equation}
  \label{eq:approx-radius-solution}
  r(\tau) = r_0\rkl{1-\tau^2}^{a/3}.
\end{equation}
We note that free-fall proceeds very slowly at the beginning. After 50\% of a freefall time the radius has decreased by only 16\%, and after 99\% of a freefall time the radius is only an order of magnitude smaller. Assuming the mass in the sphere, $M=4\pi/3\,\rho(t)r(t)^3$, to be constant during the collapse, equation~\ref{eq:approx-radius-solution} gives
\begin{equation}
  \label{eq:approx-density-solution}
  \rho(\tau)=\rho_0\rkl{1-\tau^2}^{-a}.
\end{equation}
With this equation, we can now explicitly compute the density as a function of time starting with an initial density $\rho_0$. We note that a sphere with initially uniform density maintains the uniform density distribution during the entire free-fall collapse. In addition, the time derivative of the density, which we need later, is given by
\begin{equation}
  \label{eq:density-time-derivative}
  \frac{\dif \rho}{\dif t} = \frac{12Q}{a}\,\rho_0^2 t\, \rkl{1-Q\rho_0t^2}^{-a-1},
\end{equation}
where we introduce
\begin{equation}
  Q=\frac{32G}{3\pi}
\end{equation}
to simplify many of the equations.

\section{Time evolution of the density PDF}%

\subsection{Analytic description}%
\label{sec:analytic-description}

The density PDF gives a statistical description of the distribution of density in a system. In the previous section we described a single sphere with uniform density distribution, whose PDF is a Dirac $\delta$ function, $\pdf(\rho',\tau)=\delta(\rho'-\rho(\tau))$. In the following we therefore consider a large ensemble of uniform density spheres with different initial densities $\rho_0(\tau=0)$.We must distinguish between the (volume-weighted) density PDF and its corresponding mass-weighted distribution,
\begin{equation}
  \pdfv = \frac{1}{V_\mathrm{tot}} \frac{\dif V}{\dif\rho},
\end{equation}
\begin{equation}
  \pdfm = \frac{1}{M_\mathrm{tot}} \frac{\dif M}{\dif\rho}.
\end{equation}
The distributions are normalised to the total volume, $V_\mathrm{tot}$, and the total mass, $M_\mathrm{tot}$, respectively, such that the integral over the PDF is unity, $\int_{0}^{+\infty} \pdf\, \dif\rho\!=\!1$.

\begin{figure}
  \centering
  \includegraphics[width=8cm]{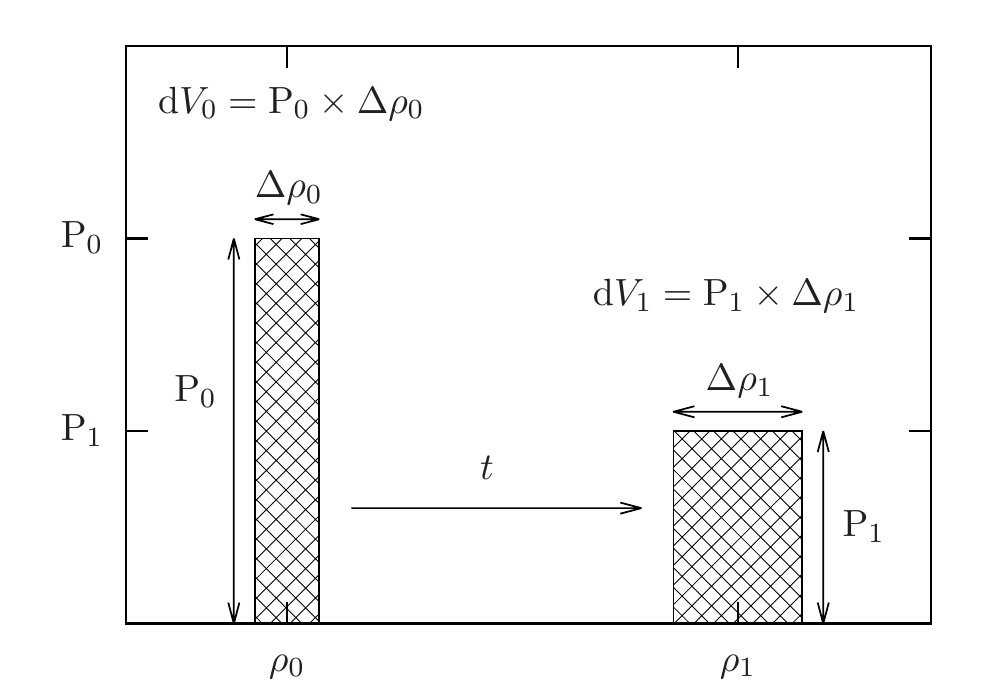}
  \caption{Sketch of the time evolution of a density bin in the PDF. The conservation of probability translates into the conservation of fractional volume.}
  \label{fig:pdf-ff-time-evol-sketch}
\end{figure}

For free-fall from rest, the density is a monotonic function of time such that, if we consider two separately collapsing uniform spheres having initial densities $\rho_i(0)$ and $\rho_j(0)\,(>\rho_i(0))$, then for all times $t$ satisfying $0\leq t\leq t_{\rm ff}(\rho_j(0))$, $\rho_i(t)<\rho_j(t)$, i.e. the evolutionary paths, $\rho_i(t)$ and $\rho_j(t)$ do not cross. We can use this mathematical property to derive an analytic description for the time evolution of the density PDF. For a (large) set of independently collapsing spheres with different initial densities we obtain an initial density PDF, for which we can use the discretisation illustrated on figure~\ref{fig:pdf-ff-time-evol-sketch}. Due to the conservation of probability density, the area of a bin on this plot stays constant over time, and so
\begin{equation}
  \label{eq:vw-pdf-time-evolution}
  \pdfv(t_1) = \pdfv(t_0) \frac{\Delta\rho_0}{\Delta\rho_1}.
\end{equation}
Starting with an arbitrary but fixed $\Delta\rho_0$ at time $t_0$, the value for $\Delta\rho_1$ at time $t_1$ can be calculated from the analytic function for $\rho(t)$, as demonstrated below.

We can transform $\pdfv$ to $\pdfm$ using
\begin{equation}
  \label{eq:conversion-mass-to-volume-weighting}
  \frac{\dif V}{\dif\rho} = \frac{\dif V}{\dif M}\frac{\dif M}{\dif\rho} = \frac{1}{\rho}\frac{\dif M}{\dif\rho}.
\end{equation}
Combining this with equation~\ref{eq:vw-pdf-time-evolution}, we obtain
\begin{equation}
  \label{eq:mw-pdf-time-evolution}
  \pdfm(t_1) = \pdfm(t_0) \frac{\rho_1}{\rho_0} \frac{\Delta\rho_0}{\Delta\rho_1}.
\end{equation}

The bin width $\Delta\rho_1$ can be written in terms of $\Delta\rho_0$, $\rho_0$ and $t$, using equation~\ref{eq:approx-density-solution},
\begin{eqnarray}\nonumber
  \Delta\rho_1&=&(\rho_0\!+\!0.5\Delta\rho_0)\!\rkl{1\!-\!Q(\rho_0\!+\!0.5\Delta\rho_0)t^2}^{\!-a}\\
  &&-(\rho_0\!-\!0.5\Delta\rho_0)\!\rkl{1\!-\!Q(\rho_0\!-\!0.5\Delta\rho_0)t^2}^{\!-a}\!.\hspace{0.8cm}
\end{eqnarray}
In the limit as $\Delta\rho_0\rightarrow0$, we have
\begin{equation}
  \label{eq:time-evol-factor-q}
  q\equiv\lim_{\Delta\rho_0\rightarrow0}\left\{\frac{\Delta\rho_0}{\Delta\rho_1}\right\} = \frac{\rkl{1-\tau^2}^{a+1}}{1+\tau^2\rkl{a-1}},
\end{equation}
where we have defined $q$ for convenience in subsequent analyses.

\subsection{Evolution of the high-density part of the PDF}%

We now make use of the simple approximation for $\rho(t)$ (equation~\ref{eq:approx-ff-solution}), to derive the time evolution of the density PDF and, in particular, we focus on the high-density part. In this context, it is convenient to use logarithmic scaling. Specifically we need to determine
\begin{equation}
  \label{eq:pdf-time-evol-split}
  \frac{\dif \ln \pdf}{\dif \ln \rho} = \frac{\dif \ln \pdf}{\dif \rho_0} \frac{\dif \rho_0}{\dif \ln \rho},
\end{equation}
as a function of density, $\rho$, and time, $t$. The high-density slope of the PDF is then given by the limit as $\tau\rightarrow1$. The second term on the right-hand side of equation~\ref{eq:pdf-time-evol-split} is
\begin{equation}
  \frac{\dif\rho_0}{\dif \ln \rho} = \rho_0\left(1 + \dfrac{a \tau^2}{1-\tau^2}\right)^{-1}
\end{equation}
and vanishes as $\tau\rightarrow1$, because the second term in the brackets dominates. For the first term on the right-hand side of equation~\ref{eq:pdf-time-evol-split} we need to distinguish between $\pdfv$ and $\pdfm$. Using equations~\ref{eq:vw-pdf-time-evolution} and \ref{eq:mw-pdf-time-evolution} the derivative for the first case takes the form
\begin{equation}
  \label{eq:pdf-vw-split}
  \frac{\dif\ln \pdfv(t_1)}{\dif\rho_0} = \frac{\dif\ln \pdfv(t_0)}{\dif\rho_0} + \frac{\dif\ln q}{\dif\rho_0},
\end{equation}
and for the latter one it takes the form
\begin{eqnarray}\nonumber
\frac{\dif\ln \pdfm(t_1)}{\dif\rho_0}&=&\frac{\dif\ln \pdfm(t_0)}{\dif\rho_0}\\\label{eq:pdf-mw-split}
&&\hspace{0.3cm}+\frac{\dif\ln q}{\dif\rho_0}+\frac{\dif\ln \rho/\rho_0}{\dif\rho_0},\hspace{0.5cm}.
\end{eqnarray}
The term containing $q$ in equations~\ref{eq:pdf-vw-split} and \ref{eq:pdf-mw-split} reads
\begin{equation}
\frac{\dif\ln q}{\dif \rho_0}=-\frac{\tau^2}{\rho_0}\rkl{\frac{a+1}{1-\tau^2} + \frac{a-1}{1+\tau^2 \rkl{a-1}}}
\end{equation}
and diverges as $\tau\rightarrow1$, because of the first term in the brackets. This means that at late times ($\tau\rightarrow1$) the right-hand sides of equations~\ref{eq:pdf-vw-split} and \ref{eq:pdf-mw-split} are both dominated by the second term, $\dif\ln q/\dif\rho_0$. Consequently, as long as it is finite, the slope of the initial PDF has very little influence on the late-time evolution of the high-density tail. Finally, the last term in equation~\ref{eq:pdf-mw-split} is given by
\begin{equation}
  \frac{\dif\ln \rho/\rho_0}{\dif\rho_0} = -\dfrac{a \tau^2}{\rho_0 (1-\tau^2)}.
\end{equation}
The slope of the high-density tail of the PDF at time $t$ is calculated by multiplying the individual terms and taking the limit for $\tau\rightarrow1$. As long as the slope of the initial PDF is finite, this yields slopes
\begin{equation}
  \label{eq:pl-slope-vol-weighted}
  \lim_{\tau\rightarrow1}\frac{\dif \ln \pdfv}{\dif \ln \rho} = -\frac{1}{a}-1,
\end{equation}
\begin{equation}
  \label{eq:pl-slope-mass-weighted}
  \lim_{\tau\rightarrow1}\frac{\dif \ln \pdfm}{\dif \ln \rho} = -\frac{1}{a}.
\end{equation}
Hence, at late times the tail of the PDF has a universal slope, independent of the slope of the initial PDF, provided this is finite. Substituting $a=1.8614$, the slopes of the tail are $-1.54$ for $\pdfv$, and $-0.54$ for $\pdfm$.

It is remarkable that the slope of the initial PDF does not influence the slope of the tail at late times. This can be understood by examining the evolution of a discretised bin of the PDF (see figure~\ref{fig:pdf-ff-time-evol-sketch}). Moving from low to high densities, the width of the bin is stretched. Rewriting equation~\ref{eq:time-evol-factor-q} as a function of density contrast $\rho/\rho_0$ (see equation~\ref{eq:approx-density-solution}), and computing the limit for large density contrast, we obtain $q^{-1}\propto(\rho/\rho_0)^{1/a+1}\approx(\rho/\rho_0)^{1.54}$. This means that the initial slope is stretched over $\approx1.5$ orders of magnitude for every order of magnitude that the density contrast increases. Thus, initial features of the PDF are stretched out and become negligible by the time the gas reaches high density. What remains is then the result of \emph{how} bins are stretched. This stretching is given solely by the intrinsic properties of gravity, encrypted in the parameter $a$.

\begin{figure*}
  \begin{minipage}{\textwidth}
  \centering
  \includegraphics[width=8cm]{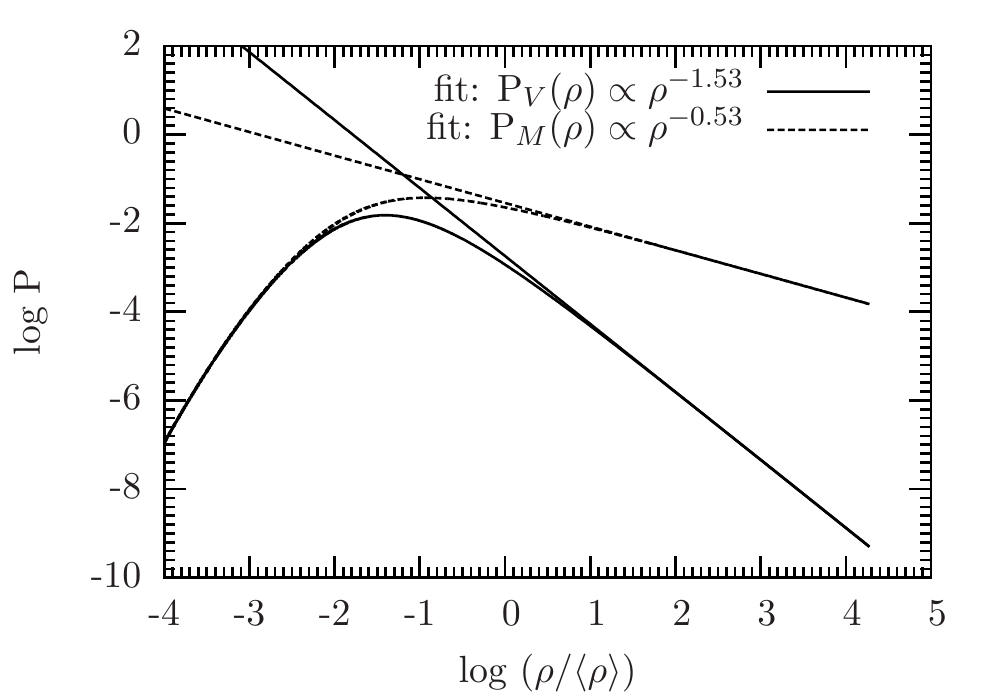}
  \includegraphics[width=8cm]{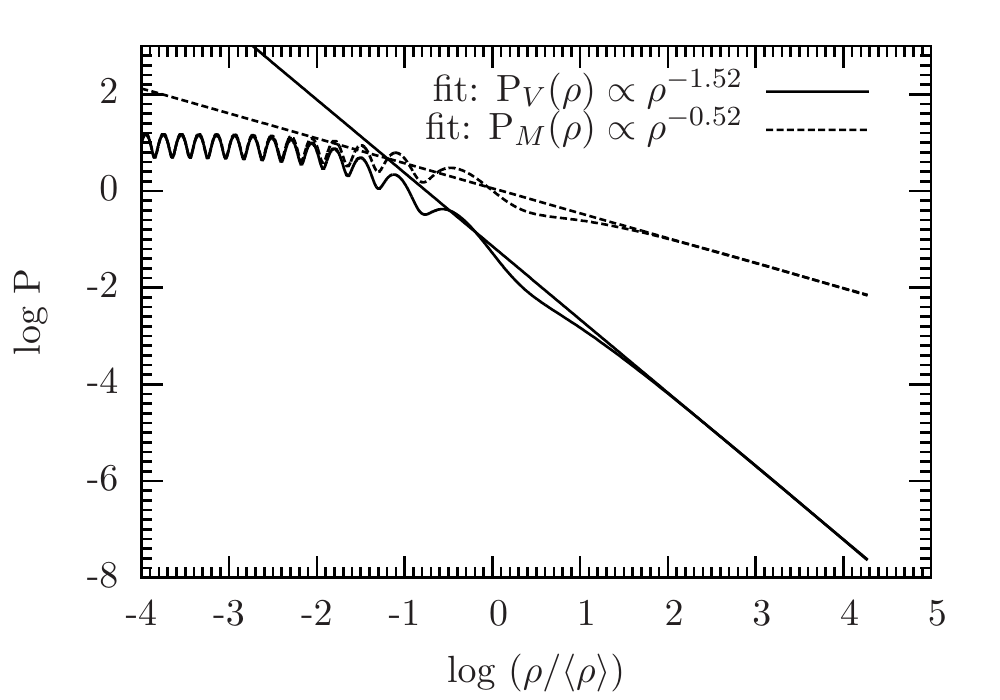}
  \caption{Evolved density PDFs. The PDF in the left plot started as a log-normal. The PDF in the right plot started as a sine wave. Independent of the initial functional form, the PDFs develop a power-law tail at high densities. The slopes of the fits are in good agreement with the analytic solution.}
  \label{fig:example-pdf-evolution-high-dens-tail}
  \end{minipage}
\end{figure*}

\begin{figure}
  \centering
  \includegraphics[width=8cm]{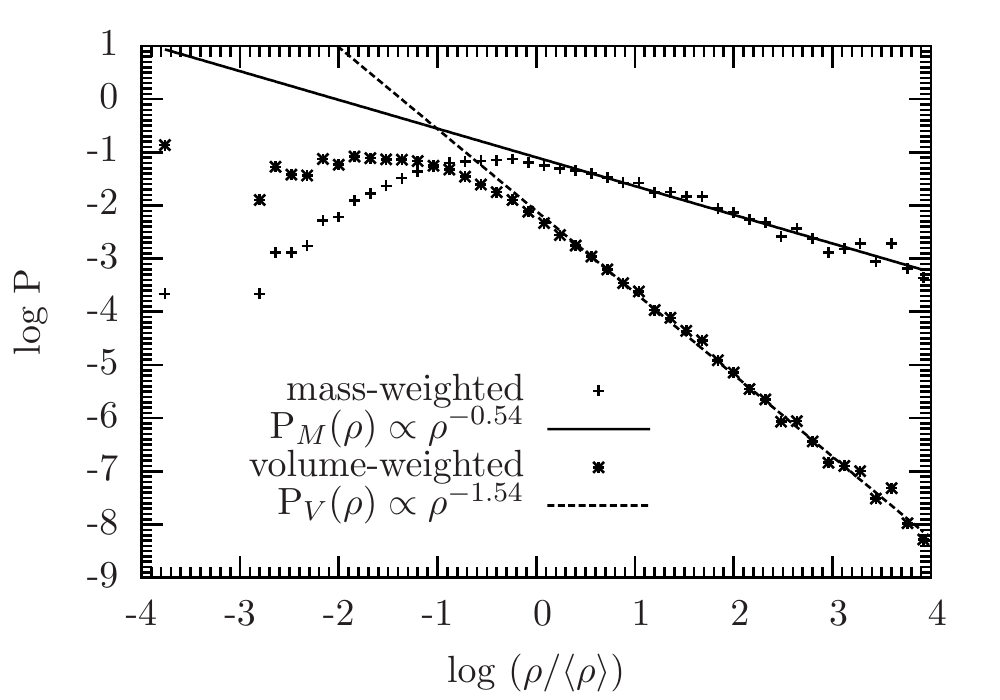}
  \caption{Density PDFs for the Monte-Carlo simulation of an ensemble of independent collapsing spheres. The slopes of the fits are in good agreement with the analytic solution.}
  \label{fig:monte-carlo-high-dens-slopes}
\end{figure}

We illustrate this in figure~\ref{fig:example-pdf-evolution-high-dens-tail}, where we plot late-time density PDFs for two different initial functions, one starting with a log-normal PDF, and one with a sinusoidal PDF, where we followed the evolution according to our approximate relations (equations~\ref{eq:approx-radius-solution} and~\ref{eq:approx-density-solution}). In addition, we perform a Monte-Carlo simulation of independent collapsing spheres. We set up an ensemble of spheres with initial radii $r_i(t=0)$ and densities $\rho_i(t=0)$ such that the combined density PDF is log-normal. The number of spheres is varied from $10^4$ to $10^7$ per ensemble. We then let the spheres start to collapse simultaneously and follow the time evolution of the radius according to equation~\ref{eq:approx-radius-solution} and the density according to equation~\ref{eq:approx-density-solution}. Spheres with higher initial density collapse quickly leading to the formation of the power-law tail as expected. Once a sphere $j$ with initial density $\rho_{j,0}\equiv\rho_j(t=0)$ has collapsed entirely ($t_\mathrm{sim}\ge t_\mathrm{ff}(\rho_{j,0})$), we exclude it from the ensemble. Figure~\ref{fig:monte-carlo-high-dens-slopes} shows the measured PDFs after $\sim10\%$ of the gas is collapsed. After an initial formation phase (see figure~\ref{fig:lognormal-tail-development}) the tail of the PDFs shows the expected slope until the majority of the spheres have collapsed and the distribution is dominated by fluctuation effects due to small number statistics.

\subsection{Link between analytic model and turbulent unstable gas}%

Up to this point, we only considered the spherically symmetric collapse of a sphere with uniform density and the collapse of an ensemble of spheres in the Monte-Carlo simulation, respectively. However, both real as well as simulated systems of gravitationally unstable gas are more complicated. There are many physical aspects that influence the gaseous region such as turbulence, self-gravity, magnetic fields, chemical, or thermal effects. As the main assumption of our model is the condition of free-fall, we need to discuss whether this condition is fulfilled in real and simulated systems, respectively.

\begin{figure}
  \centering
  \includegraphics[width=8cm]{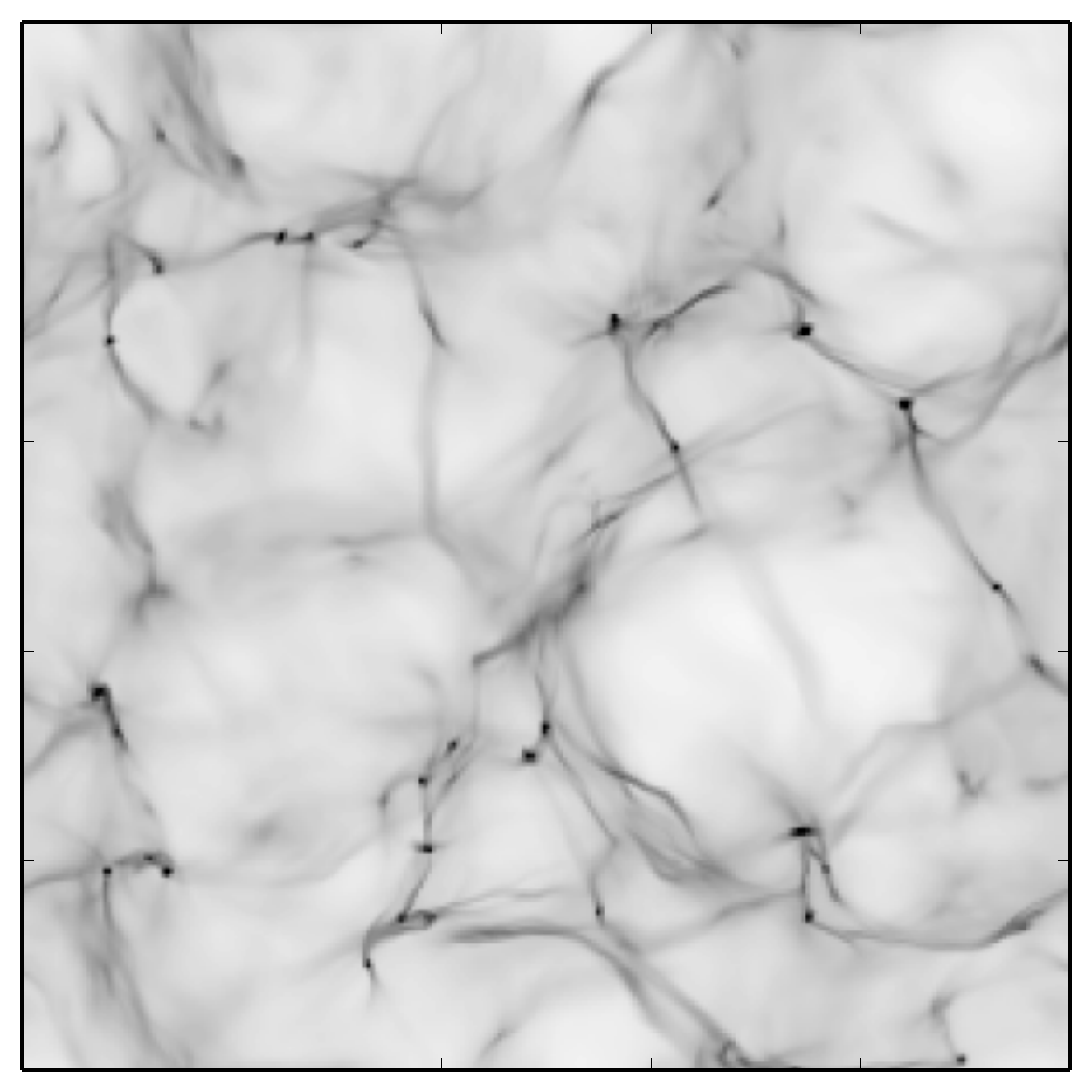}
  \caption{Column density plot from an isothermal hydrodynamical simulation of turbulent self-gravitating gas. The turbulence creates a hierarchy of structures, where the densest regions condense into gravitationally unstable cores. The details of the structure depend on the properties of the turbulence and the gas but qualitatively the effects of self-gravitating turbulent cores looks very similar for a large range of parameters.}
  \label{fig:hydro-turblence-with-cores}
\end{figure}

On of the most important processes in the fragmentation of a gaseous region is turbulence \citep{Elmegreen04,Scalo04,MacLow04,HennebelleFalgarone2012}. Turbulent motions create a hierarchy of structures that break up ordered systems into smaller subsystems. Turbulent regions are therefore not spherically symmetric and do not have uniform density distributions as a whole. Rather, they form elongated structures with local overdensities as demonstrated by figure~\ref{fig:hydro-turblence-with-cores}, a column density snapshot from a hydrodynamical simulation of self-gravitating turbulent gas. The large-scale supersonic motions break up and cascade down to transsonic motions in the dense cores, which is confirmed by theory as well as observations \citep[e.g.,][]{Larson81,BerginTafalla2007}. Once the densest cores start collapsing by passing the Jeans instability (see appendix~\ref{sec:thermodynamics}), the collapse from cores down to stars, i.e., the star formation process, occurs within a free-fall time. Although the individual fragmentation of cores is still a matter of debate, there is a consensus that the collapse time scale in real and simulated systems is the free-fall time \citep[e.g.,][]{Krumholz05,PadoanEtAl1997,HennebelleChabrier2008,ShadmehriElmegreen2011,HennebelleFalgarone2012}. Thus, it seems appropriate to apply a free-fall model.

Apart from the fact that cores collapse in free-fall one can ask whether the fragments of a collapsing sphere can again be described by a sphere, i.e., whether the fragmentation in this idealised picture is self-similar. \citet{Hunter1962} found that perturbations in the density structure of a uniform sphere lead to fragmentation of the sphere and that all perturbing wavelengths grow equally. The resulting fragmented structure will thus depend on the Fourier transform of the perturbation field. If one assumes perturbations of equal amplitude in random directions, the resulting pattern in real space reveals clumpy roundish structures. \citet{Falle1972} extended the work by \citet{Hunter1962} and investigated the fragmentation of a uniform spheroid. They conclude from a linear perturbation analysis that the spheroid most likely fragments by slicing parallel to the axis. The lack of spherical symmetry also leads to the conclusion that the dominant perturbations are of the order of the Jeans length or larger, resulting in a small number of large fragments rather than in a large number of small fragments. All together it seems reasonable to assume that the collapsing and fragmenting sphere breaks up into large clumpy objects that are unstable themselves, i.e., that the fragmentation can be assumed roughly self-similar.

\subsection{Comparison to hydrodynamic simulations}%

The slope of the tail of the PDF has been measured in numerical simulations of turbulent self-gravitating clouds. In the high-density collapsing regime the PDF is observed to subscribe to a power-law, $\mathrm{PDF}(\rho)\propto\rho^{\,\mu}$. The slope was measured for the first time by \citet{SlyzEtAl2005} who found $\mu=-1.5$, which agrees very well with our analytic prediction. High-resolution simulations by \citet{KritsukEtAl2011} reveal a slope of $\mu=-1.67$ at intermediate to high densities, and $\mu=-1.5$ at very high densities. Recently, \citet{CollinsEtAl2011} have performed simulations of magnetised clouds, and find slopes ranging from $\mu=-1.64$ to $\mu=-1.80$, steeper than in the purely hydrodynamic case; the stronger the magnetic field, the steeper is the tail of the PDF.

\begin{figure*}
  \begin{minipage}{\textwidth}
    \centering
    \includegraphics[width=8cm]{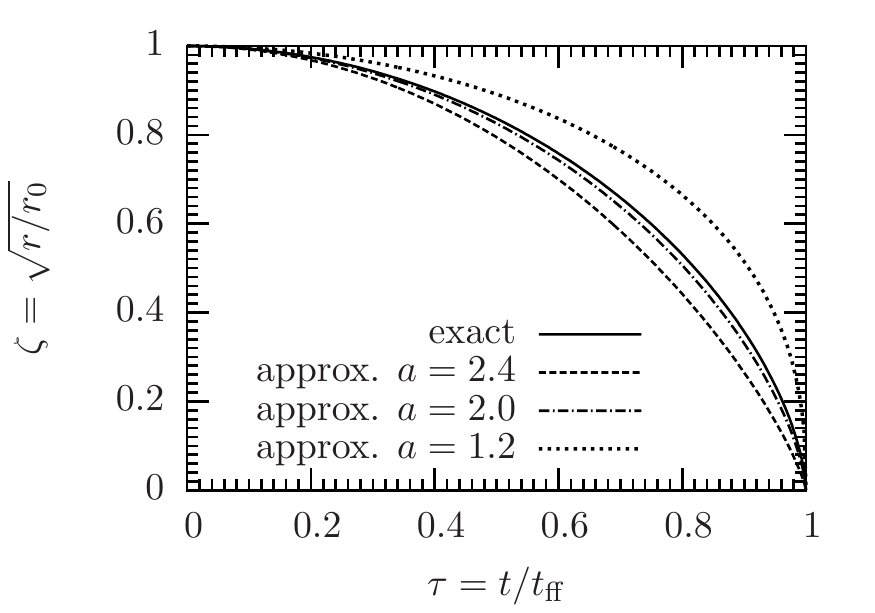}
    \includegraphics[width=8cm]{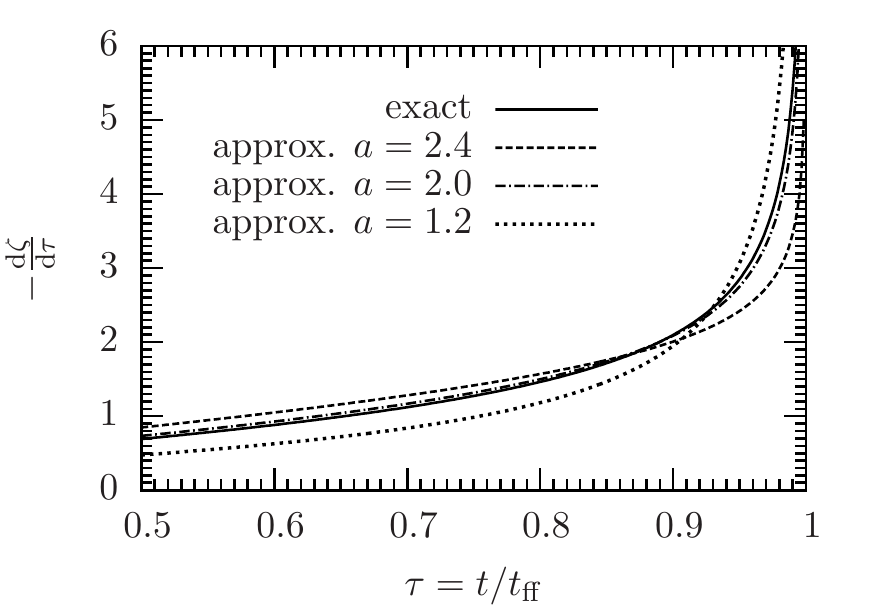}
  \end{minipage}
  \caption{{\sc Left.} Dimensionless radius against dimensionless time, for different values of $a$. {\sc Right.} Corresponding plots of dimensionless inward radial velocity against dimensionless time. Large values for $a$ mimic a delayed collapse that speeds up towards the end. Small values of $a$ accelerate the collapse early on.}
  \label{fig:collapse-radius-different-a}
\end{figure*}

We can understand this behaviour qualitatively by investigating the influence of the parameter $a$. Although $a$ is introduced as a simple mathematical device, to match the approximate function (equation~\ref{eq:approx-ff-solution}) to the true collapse solution, we can mimic other physical effects by changing its value. Figure~\ref{fig:collapse-radius-different-a} shows the radius of the collapsing sphere, and its time derivative, as a function of time for different values of $a$. It would be ideal to have a complete description and derivation of the slope of the PDF as a function of initial collapse velocities incorporated in the initial differential equation~\ref{eq:free-fall-DGL}. However, the simple approximation and the subsequent derivative of the slope of the PDF is not possible any more in an analytic way. We therefore stick to this crude approximation to Newton's law.

For $a<1.8614$, the collapse is delayed initially, but speeds up towards the end. This is qualitatively what happens in marginally unstable density regimes, where the early phase of the collapse is still influenced by stabilising effects like thermal pressure or magnetic fields. At later stages of the collapse, gravity becomes increasingly dominant. Hence, structures with different densities are influenced differently by stabilising effects. The collapse of gas with initial density $\rho_i$ is more strongly delayed than the collapse of gas with initial density $\rho_j>\rho_i$. This results in a larger stretching of the density bin $\Delta\rho = \rho_j- \rho_i$, compared with the case of free-fall. The larger the stabilising effects, e.g., the magnetic field strength \citep{CollinsEtAl2012}, the larger is the stretching and the steeper the tail.

Conversely, the tail is shallower if collapse is faster than pure free-fall. This is the case in converging flows, where the collapse does not start with zero velocity, but instead with an initial converging velocity field \citep{Ballesteros03, Klessen05, Banerjee09}.

\section{Connection between the PDF and mass accretion}%

\subsection{Derivation of the accretion rate}%

The mass accretion rate is defined as the rate at which mass evolves above a threshold density $\rho_\mathrm{accr}$. In the free-fall approximation this is
\begin{equation}\label{EQN:ACCNRATE}
\dot{M}_{\rho_\mathrm{accr}}(t) = \left.\frac{\dif M}{\dif\rho}\right|_{\rho_\mathrm{accr}}\,\left.\frac{\dif\rho}{\dif t}\right|_{\rho_\mathrm{accr}},
\end{equation}
and the problem is to evaluate the derivatives on the right side of equation~\ref{EQN:ACCNRATE}, as functions of $\rho_\mathrm{accr}$ and $t$.

Following the argument in section~\ref{sec:analytic-description} allows us to express the first term on the right hand side as a function of $\pdfm(t_1)$, where $t_1$ corresponds to the time, when the gas reached the density $\rho_\mathrm{accr}$. It is important to note that the bins of constant volume in figure~\ref{fig:pdf-ff-time-evol-sketch} do not conserve mass over time. The first term on the right side of equation~\ref{EQN:ACCNRATE} therefore needs to be adjusted to ensure mass-conservation of the system. The mass of the bin increases by a factor of $M_1/M_0$ which we have to correct for yielding
\begin{align}
  \label{eq:mw-pdf-accretion-time-evolution}
  \left.\frac{\dif M}{\dif\rho}\right|_{\rho_\mathrm{accr}} &= \frac{M_0}{M_1}\pdfm(t_1)\\
  &= \frac{\rho_0}{\rho_1}\pdfm(t_1)\\
  &= \pdfm(t_0) \frac{\Delta\rho_0}{\Delta\rho_1}.
\end{align}

In order to determine the rate at which mass evolves above $\rho_\mathrm{accr}$ at time $t$, we need to solve for the initial density $\rho_0$ that evolves to $\rho_\mathrm{accr}$ at this time, i.e. to solve $\rho(\rho_0,t)\!=\!\rho_\mathrm{accr}$ and obtain  $\rho_0(\rho_\mathrm{accr},t)$. If we set $a\!=\!1.8614$, equation~\ref{eq:approx-density-solution} can not be inverted analytically to obtain $\rho_0(\rho_\mathrm{accr},t)$. However, if we set $a\!=\!2$ (which still only deviates from the exact collapse solution by less than $4\%$, see figure~\ref{fig:error-approx-solution}), equation~\ref{eq:approx-density-solution} becomes
\begin{equation}
\rho_x(\rho_0,t) = \rho_0\rkl{1-Q\rho_0\,t^2}^{-2},
\end{equation}
and can be inverted analytically to give
\begin{equation}
\rho_0(\rho_x,t) = \frac{2Q \rho_x t^2 + 1}{2Q^2 \rho_x t^4} + \sqrt{\rkl{\frac{2Q \rho_x t^2 + 1}{2Q^2 \rho_x t^4}}^2 - \frac{1}{Q^2 t^4}}.
\end{equation}

\begin{figure}
  \centering
  \includegraphics[width=8cm]{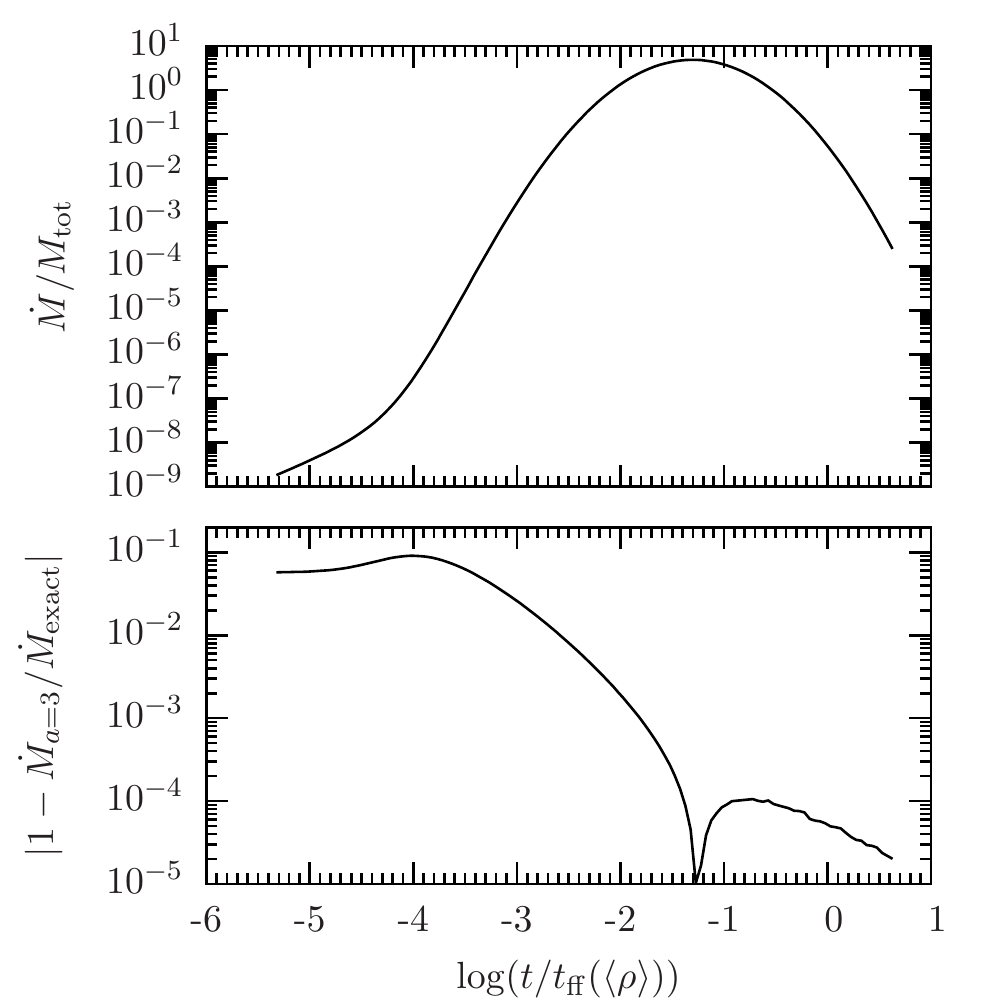}
  \caption{The accretion rate with $a\!=\!2$, and the fractional deviation between this rate and the exact solution.}
  \label{fig:accretion-plus-error}
\end{figure}

This solution for $\rho_0(\rho_x,t)$ can then be used to calculate $\pdfm$ (equation~\ref{eq:mw-pdf-accretion-time-evolution}) and the time derivative of the density (equation~\ref{eq:density-time-derivative}), for substitution into equation~\ref{EQN:ACCNRATE}. Figure~\ref{fig:accretion-plus-error} shows the resulting accretion rate, and the deviation arising from using the approximate solution (equation~\ref{eq:approx-ff-solution}) with $a=2$, in place of the exact solution (equation~\ref{eq:exact-ff-solution}). The dip in the lower plot is due to the switch from a positive to a negative deviation.

\subsection{Log-normal density PDFs}%

Although according to our approximation the high density slope is unaffected by the functional details of the initial PDF, we need to adopt an initial functional form of the PDF in order to apply the free-fall model to the accretion rate of a collapsing system.

Isothermal non-self-gravitating gas develops a log-normal PDF in response to driven supersonic turbulence (see Introduction). Although astrophysical systems are often not isothermal and are always self-gravitating, a significant fraction of the interstellar medium is dominated by kinetic motions. Under this circumstance, thermal effects and self-gravity play a minor role, and the density PDF approximates to a log-normal form \citep{MacLow04,Elmegreen04,McKeeOstriker2007,GoodmanEtAl2009,KainulainenEtAl2009}. This overall highly dynamical state of clouds and cloud complexes seems to be in contrast to our model of collapsing spheres from rest. However, the large scale turbulent motions cascade down to transsonic or even subsonic motions, such that many of the dense cores that populate the high-density extreme of the PDF appear to start their collapse from relatively quiescent initial conditions, in which the turbulence is at best transsonic \citep[e.g.,][]{BerginTafalla2007}. Our assumption of a negligible net effect of the initial turbulence on the contraction of the spheres is clearly a deficiency that needs to be extended in a future work with a more general theory. However, as it is unclear to what extent the turbulence results in a net initial contraction or expansion, we apply our model to the log-normal density PDF, viz.
\begin{equation}
  \pdf(\rho) \propto \exp\rkl{-\frac{(\ln(\rho/\!\skl{\rho}) - \ln(\rho_\mathrm{peak}/\!\skl{\rho})^2}{2\sigma^2}},
\end{equation}
with variance
\begin{eqnarray}\label{eq:turbulent-pdf-width:1}
\sigma^2&=&\ln(1+b^2\mathcal{M}^2)\\\label{eq:turbulent-pdf-width:2}
&=&\ln(1+3\mathcal{M}_\mathrm{comp}^\gamma),
\end{eqnarray}
and peak density
\begin{equation}
  \label{eq:turbulent-pdf-peak}
  \ln\rkl{\frac{\rho_\mathrm{peak}}{\skl{\rho}}} = -\frac{\sigma^2}{2}.
\end{equation}

The variance, $\sigma^2$, can be related to the dynamics of the system in two different ways. In the first one (see equation~\ref{eq:turbulent-pdf-width:1}), the empirical parameter $b$ depends on the nature of the turbulence. \citet{Federrath10b} have  investigated driven turbulence, and find $b=1/3$ for purely solenoidal driving and $b=1$ for purely compressive driving. Here, the quantity $\mathcal{M}=v_\mathrm{rms}/c_\mathrm{s}$ is the rms Mach number of the turbulence. In the second one (see equation~\ref{eq:turbulent-pdf-width:2}), the best fit value for the empirical parameter $\gamma$ is $\gamma\!=\!1.7$ and $\mathcal{M}_\mathrm{comp}=v_\mathrm{comp,rms}/c_\mathrm{s}$ is the Mach number of the compressive motions \citep{KonstandinEtAl2012b}. To begin with, we set $b$ to unity and vary the dynamics via the Mach number, $\mathcal{M}$. We use normalised units with respect to the average density $\skl{\rho}$.

\begin{figure}
  \centering
  \includegraphics[width=8cm]{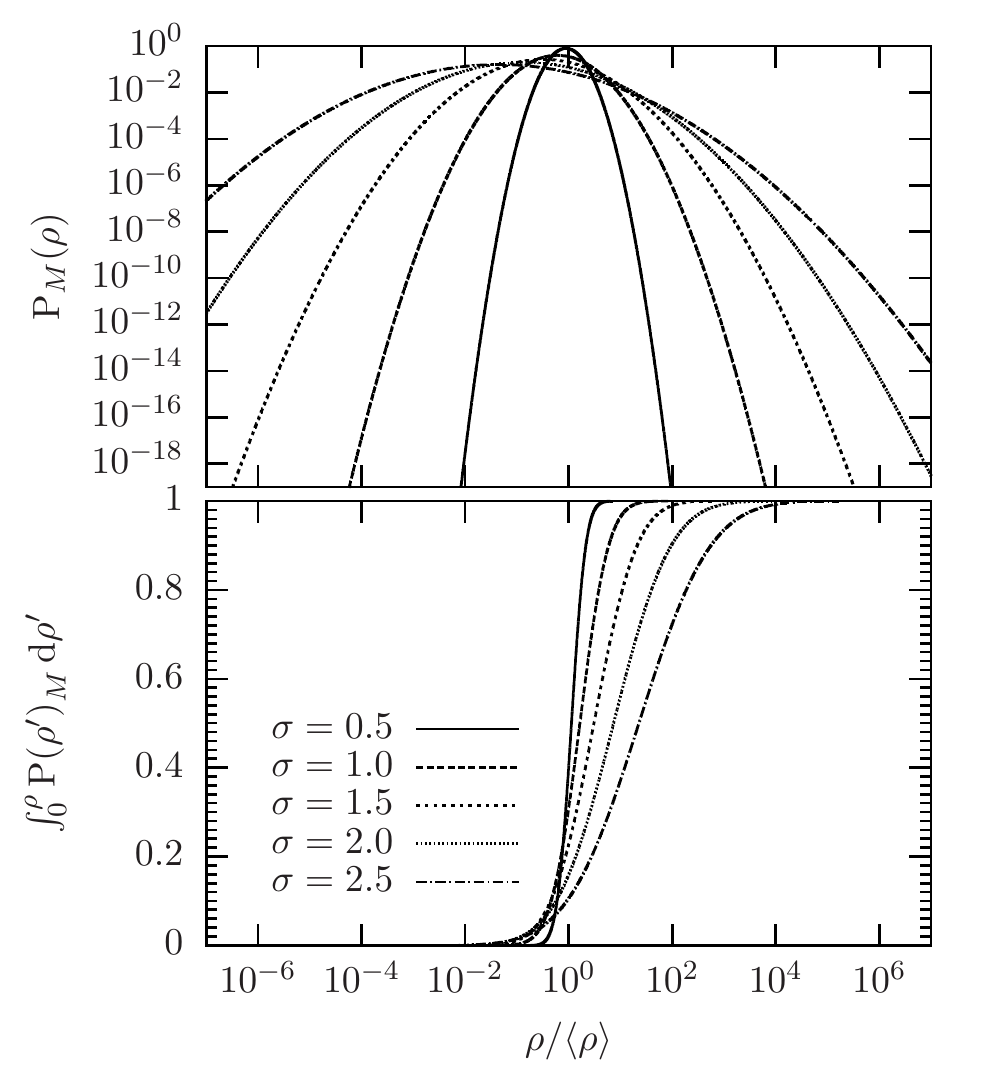}
  \caption{Shown are $\pdfm$ and cummulative mass distribution, for different standard deviations, $\sigma$. The mass distribution is highly asymmetric.}
  \label{fig:pdf0-mass-distribution}
\end{figure}

Figure~\ref{fig:pdf0-mass-distribution} illustrates $\pdfm$, and cummulative distributions ($\int_0^{\rho}\pdf(\rho')\,\dif\rho'$), for different values of the standard deviation, $\sigma$. The mass distributions of the log-normal PDFs are highly asymmetric, in the sense that there is very little mass at densities $\rho<\rho_\mathrm{peak}$ for large $\sigma$. Table~\ref{tab:lognormal-mass-distribution-sigma} records the peak density, the fraction of mass below the peak, and the density $\rho_{50\%}$ at which the integral $\int_0^{\rho}\pdf(\rho')\,\dif\rho'\!=\!0.5$. These numbers are critical to understanding how the accretion rate varies with time.

\begin{table}
  \caption{Key properties of initial log-normal density PDFs}
  \begin{center}
  \begin{tabular}{clcc}
    $\sigma$ & $\ln\rkl{\dfrac{\rho_\mathrm{peak}}{\skl{\rho}}}$ & $\dfrac{M(\rho<\rho_\mathrm{peak})}{M_\mathrm{tot}}$ & $\ln\rkl{\dfrac{\rho_\mathrm{50\%}}{\skl{\rho}}}$\\
    \hline
    0.5 & $-0.125$ & 0.309 & $-2.9\phantom{0}$ \\
    1.0 & $-0.5$   & 0.157 & $-1.5\phantom{0}$ \\
    1.5 & $-1.125$ & 0.067 & $-0.72$ \\
    2.0 & $-2.0$   & 0.023 & $-0.14$ \\
    2.5 & $-3.125$ & 0.006 & $\phantom{-}0.31$ \\
    \hline
  \end{tabular}
  \end{center}

Column 1 gives values of the standard deviation, $\sigma$. Column 2 gives the density at the peak, $\rho_\mathrm{peak}$, in units of $\skl{\rho}$. Column 3 gives the mass fraction below the peak. Column 4 gives the median density, $\rho_{50\%}$, above and below which there is equal mass. \\
  \label{tab:lognormal-mass-distribution-sigma}
\end{table}

One limitation of the model developed here is that we invoke, ab initio, a cloud having a log-normal density PDF, corresponding to a cloud without self-gravity, and we do not specify how such a cloud might have been assembled. This cloud is then assumed to undergo free-fall collapse. As a result, the time evolution of the PDF, especially at early times, may show artificial features that should not be expected, either in observations, or in numerical simulations, i.e., features that would not arise if the cloud were assembled self-consistently in the presence of self-gravity. How strong these features are depends on the density regime we are investigating. Very dense regions with $\rho\gg\skl{\rho}$ will collapse on very short time scales compared to the global characteristic time scale of the cloud, transforming the very steep fall-off of the log-normal density PDF into the relatively slow fall-off of the power-law tail. In contrast, gas at densities far below the peak, $\rho\ll\skl{\rho}$, will not be affected, because the free-fall time of this gas is very long compared to the evolutionary time scales with which we are concerned.

\begin{figure}
  \centering
  \includegraphics[width=8cm]{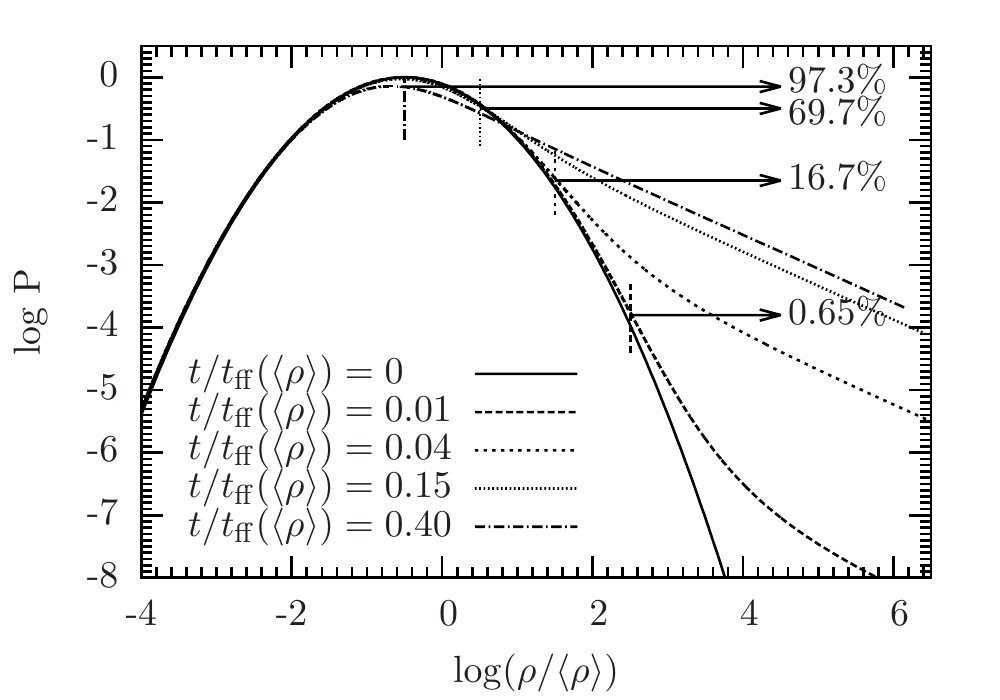}
  \caption{Formation of the power-law tail from a log-normal distribution. After $0.01t_\mathrm{ff}(\skl{\rho})$ the tail starts to deviate from the log-normal at $10^3\skl{\rho}$ and a power-law has been established at densities $5-6$ orders of magnitude above the peak density. After $5\%$ of the peak free-fall time, a full power-law tail has been established for densities above $10^4\skl{\rho}$. The threshold densities, where the tail begins for the times shown, are indicated by small vertical lines. The corresponding total fraction of mass in the high-density tail is shown at the end of the arrow.}
  \label{fig:lognormal-tail-development}
\end{figure}

Figure~\ref{fig:lognormal-tail-development} illustrates how the power-law tail is established. We define the tail as the part of the PDF that is noticably above the approximately log-normal part (this transition is located by eye, and marked with a small vertical line on figure~\ref{fig:lognormal-tail-development}), and the power-law tail is the part that has asymptoted to the universal slope given by equation~\ref{eq:pl-slope-mass-weighted}. After $0.01t_\mathrm{ff}(\skl{\rho})$ the tail starts above $\sim\! 10^{2.5}\skl{\rho}$, and the power-law tail starts above $\sim\! 10^6\rho_\mathrm{peak}$ (this falls outside figure~\ref{fig:lognormal-tail-development}). After $0.04t_\mathrm{ff}(\skl{\rho})$, the power-law tail starts above $10^4\skl{\rho}$. The reason why the switch from a log-normal PDF to a power-law tail advances steadily to ever lower densities is the highly non-linear nature of gravitational collapse, which starts very slowly and speeds up rapidly towards the end. Consequently, the impact of free-fall on the density PDF at density $\rho_x$ can only manifest itself after $\sim\!t_\mathrm{ff}(\rho_x)$, and we should expect the power-law tail to be well-defined above a density $\rho_\mathrm{tail}(t)$ given by
\begin{equation}
\frac{\rho_\mathrm{tail}(t)}{\skl{\rho}}\sim\rkl{\frac{t}{t_\mathrm{ff}(\skl{\rho})}}^{-2}.
\end{equation}
Measurements of $\rho_\mathrm{tail}(t)$ based on figure~\ref{fig:lognormal-tail-development} are indeed well fitted by
\begin{equation}
  \frac{\rho_\mathrm{tail}(t)}{\skl{\rho}}\approx0.2\rkl{\frac{t}{t_\mathrm{ff}(\skl{\rho})}}^{-2.0}.
\end{equation}
Thus, for a given density, $\rho_x$, the tail starts to deviate from the log-normal distribution at roughly $0.45t_\mathrm{ff}(\rho_x)$, when the density has increased by $\sim\!50\%$ (see equation~\ref{eq:approx-density-solution}).

The fraction of the total mass in the tail (i.e., above the density indicated by the vertical line in figure~\ref{fig:lognormal-tail-development}) is shown at the ends of the arrows on figure~\ref{fig:lognormal-tail-development}. After $0.15t_\mathrm{ff}(\skl{\rho})$ already $\sim\!70\%$ of the mass is in the tail. Above a certain threshold density, $\rho_\mathrm{accr}$, we can infer that the gas has condensed into stars, and so the total fraction of gas above $\rho_\mathrm{accr}$ gives the star formation efficiency (SFE), and the rate at which mass evolves above $\rho_\mathrm{accr}$ gives the star formation rate (SFR).

\subsection{Dependence of the accretion on the threshold density}%

\begin{figure}
  \centering
  \includegraphics[width=8cm]{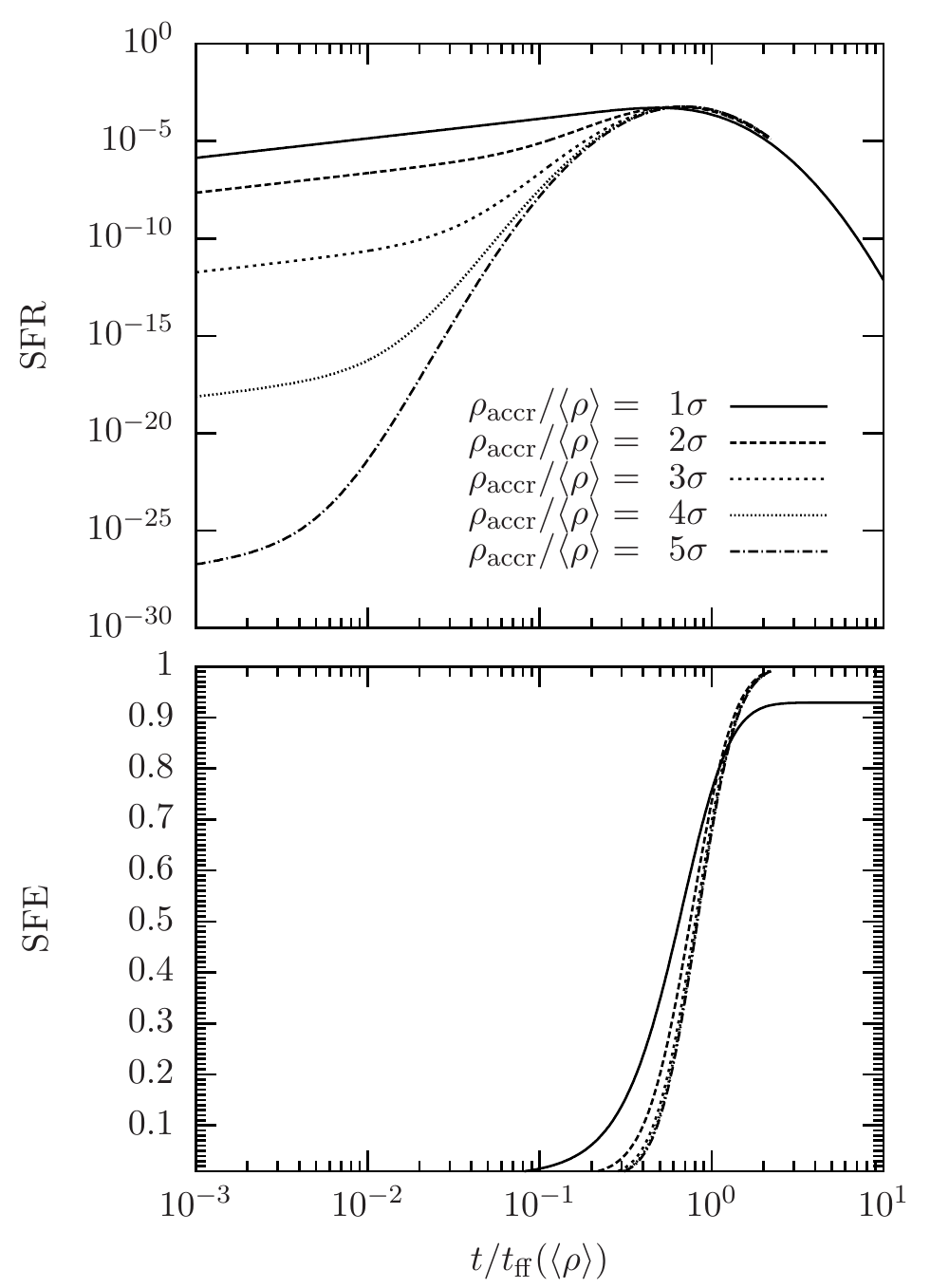}
  \caption{The star formation rate and star formation efficiency for different accretion threshold densities $\rho_\mathrm{accr}$. For $\rho_\mathrm{accr}\ge5\sigma$ the star formation efficiencies are indistinguishable.}
  \label{fig:accretion-depend-rhox}
\end{figure}

Both the SFE and the SFR depend on the threshold density, $\rho_\mathrm{accr}$. To illustrate this, we set $\mathcal{M}=b=1$ and vary the threshold density in units of $\sigma$ from $1\sigma$ to $5\sigma$. Figure~\ref{fig:accretion-depend-rhox} shows the SFR and the SFE, both as functions of time, for different threshold densities. The time axis is normalised to the free-fall time of the average density.

The initial SFR is significantly higher for lower $\rho_\mathrm{accr}$, because, if $\rho_\mathrm{accr}$ is low, there is more gas just below $\rho_\mathrm{accr}$ poised ready to evolve above $\rho_\mathrm{accr}$. However, the SFR at late times is only weakly dependent on $\rho_\mathrm{accr}$, and in all cases it peaks at $\sim\!t_\mathrm{ff}(\skl{\rho})/2$. The SFE only counts gas that passes the threshold density $\rho_\mathrm{accr}$ during the evolution, not the gas that is already located in regions with $\rho>\rho_\mathrm{accr}$ at the outset. Thus, in the case with $\rho_\mathrm{accr}/\skl{\rho}\!=\!1\sigma$, roughly 10\% of the gas is initially denser than $\rho_\mathrm{accr}$, and consequently the SFE saturates at $\mathrm{SFE}\approx0.9$.

Provided that we set $\rho_\mathrm{accr}$ sufficiently high, the SFE is very weakly dependent on $\rho_\mathrm{accr}$. Table~\ref{tab:lognormal-tail-development} records, as a function of the time, $t$, the density, $\rho_\mathrm{tail}$, at which the tail departs markedly from the log-normal portion (see figure~\ref{fig:lognormal-tail-development}), the fraction of the total mass, $m_\mathrm{tail}$, that is in the tail, and the star formation efficiencies, $\mathrm{SFE}_n$, corresponding to different $\rho_\mathrm{accr}\!=\!10^n\skl{\rho}$, for $n\!=\!5,\,6,\;{\rm and}\;7$.

\begin{table}
  \caption{Key properties of the evolution of the power-law tail}
  \begin{tabular}{cccccc}
    $\dfrac{t}{t_\mathrm{ff}(\skl{\rho})}$& $\dfrac{\rho_\mathrm{tail}}{\skl{\rho}}$&$m_\mathrm{tail}$&$\mathrm{SFE}_5$&$\mathrm{SFE}_6$&$\mathrm{SFE}_7$\\
    \hline
    0.01 & $300$ & 0.007 & $8.0\times10^{-5}$ & $4.0\times10^{-5}$ & $4.0\times10^{-5}$\\
    0.04 & $30$  & 0.17\phantom{0}           & 0.025 & 0.024 & 0.023\\
    0.15 & $3$   & 0.70\phantom{0}           & 0.35\phantom{0} & 0.35\phantom{0} & 0.35\phantom{0}\\
    0.40 & $0.3$ & 0.97\phantom{0}           & 0.78\phantom{0} & 0.78\phantom{0} & 0.78\phantom{0}\\
    \hline
  \end{tabular}

Column 1 gives the time, $t$, in units of $t_\mathrm{ff}(\skl{\rho})$. Column 2 gives the density, $\rho_\mathrm{tail}$, above which the density PDF diverges markedly from log-normal portion (i.e. the tail, see figure~\ref{fig:lognormal-tail-development}), in units of $\skl{\rho}$. Column 3 gives the mass fraction, $m_\mathrm{tail}$, above $\rho_\mathrm{tail}$. Columns 4 through 6 give the star formation efficiencies, $\mathrm{SFE}_n$, corresponding to different $\rho_\mathrm{accr}\!=\!10^n\skl{\rho}$, for $n\!=\!5,\,6,\;{\rm and}\;7$. \\
  \label{tab:lognormal-tail-development}
\end{table}

\subsection{Predicted star formation rates}\label{sec:star-formation-rates}%

\begin{figure}
  \centering
  \includegraphics[width=8cm]{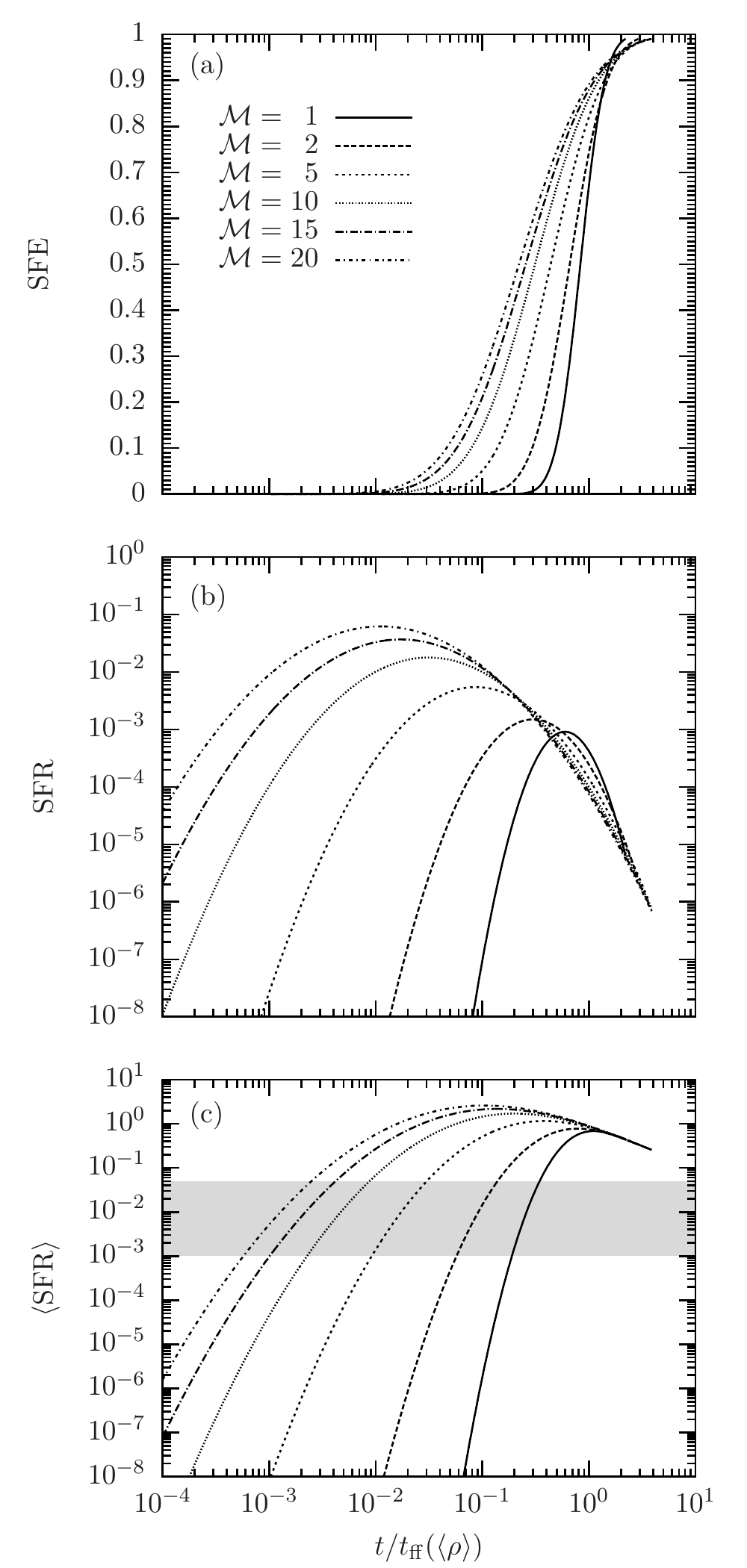}
  \caption{Star formation efficiency, normalised star formation rate, and average normalised star formation rate as a function of normalised time $t/t_\mathrm{ff}(\skl{\rho})$. The curves are plotted for different Mach numbers, i.e., for different widths of the initial PDF. The grey area in panel (c) indicates observational measurements by \citet{HeidermanEtAl2010}.}
  \label{fig:dimless-SFR}
\end{figure}

We define the normalised star formation rate, using the free-fall time at the average density and total mass, viz.
\begin{equation}
  \mathrm{SFR}(t) = \frac{\dot{M}(t)\,t_\mathrm{ff}(\skl{\rho})}{M_\mathrm{tot}},
\end{equation}
and the mean normalised star formation rate as $\skl{\mathrm{SFR}}=\mathrm{SFE}(t)/t$. Figure~\ref{fig:dimless-SFR} shows the time evolution of (a) the star formation efficiency, (b) the normalised star formation rate, and (c) the mean normalised star formation rate, for different widths of the initial density PDF. The grey shaded band indicates the range of observed average star formation rates (see next section). In more quiescent regions it takes much longer for star formation to set in, and when it finally does, the rate is very low. For highly dynamic systems with a broad density PDF, more mass is located in high-density regions. Consequently, star formation sets in much earlier and the peak star formation rates are perceptibly higher. Significant fractions of the cloud are accreted on very short time scales, so that very dynamic systems lead to a more burst-like star formation event.

\begin{figure}
  \centering
  \includegraphics[width=8cm]{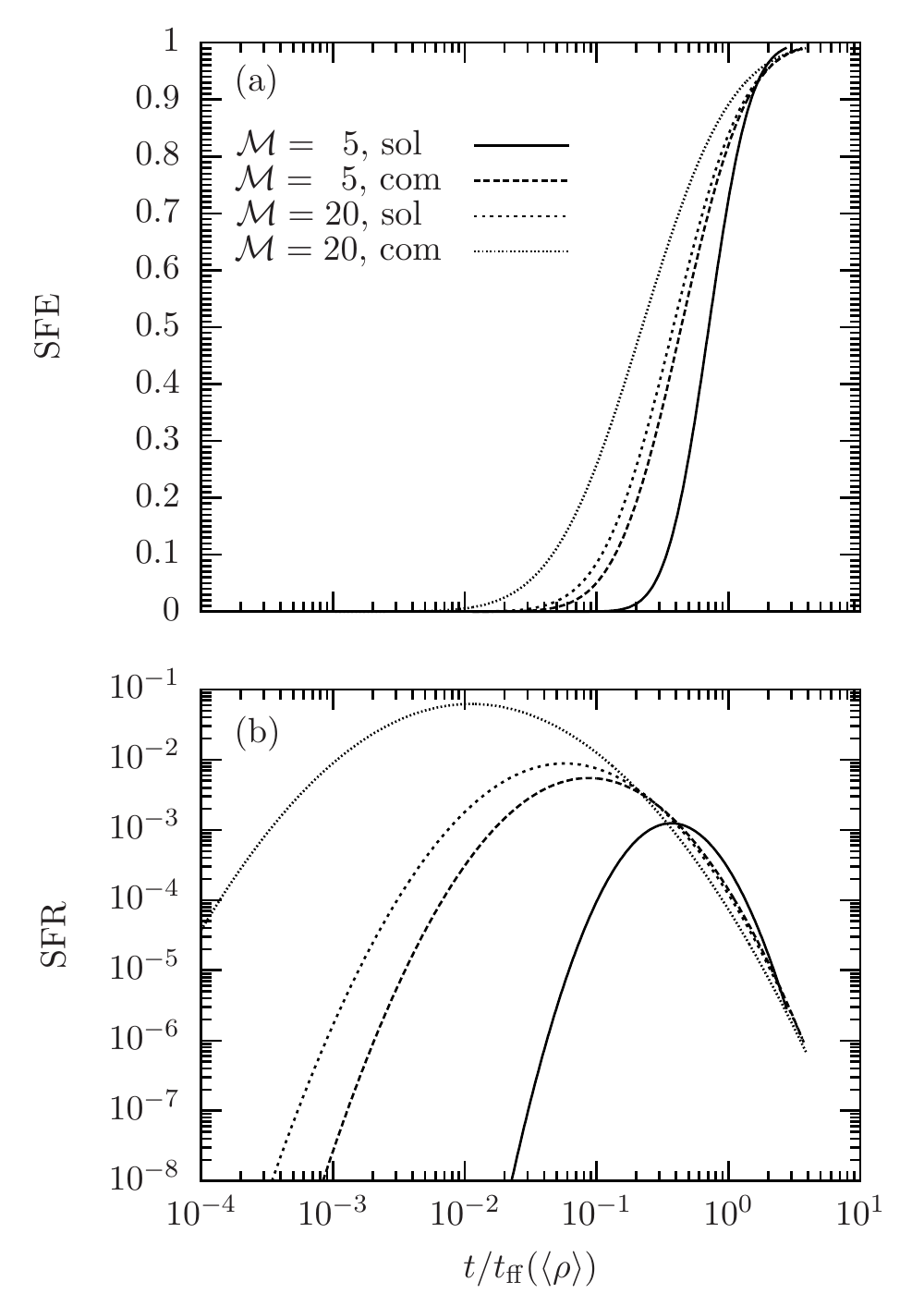}
  \caption{The star formation efficiency and normalised star formation rate, as a function of normalised time $t/t_\mathrm{ff}(\skl{\rho})$, for purely compressive and purely solenoidal driving, with $\mathcal{M}=5$ and $\mathcal{M}=20$.}
  \label{fig:dimless-SFR-sol-comp}
\end{figure}

In order to compare the star formation process for very different levels of turbulence, and different turbulent driving modes (cf., equations~\ref{eq:turbulent-pdf-width:1} and \ref{eq:turbulent-pdf-width:2}), we set the Mach number to $\mathcal{M}=5$ and $\mathcal{M}=20$, and invoke purely solenoidal driving, $b=1/3$, and purely compressive driving, $b=1$. For $\mathcal{M}=5$, this corresponds to compressive Mach numbers of $\mathcal{M}_\mathrm{comp}=0.96$ and $\mathcal{M}_\mathrm{comp}=3.48$, for pure solenoidal and pure compressive driving respectively. For $\mathcal{M}=20$ the corresponding numbers are $\mathcal{M}_\mathrm{comp}=4.88$ and $\mathcal{M}_\mathrm{comp}=17.8$. Figure~\ref{fig:dimless-SFR-sol-comp} shows the star formation efficiency, and the normalised star formation rate for all four cases. As expected, a higher compressive Mach number leads to a broader density PDF, faster star formation, and a higher star formation rate.

\section{Comparison to observations}%

Among the quantities we compute from our model, two are in theory suitable for comparing with observations, namely the density PDF, in particular the tail, and the star formation rate as a function of time, SFR(t). However, SFR(t) is in practice impossible to measure because the time evolution of molecular cloud cores is too slow to be observed. Instead we have to compare the normalised mean star formation rate, $\skl{\mathrm{SFR}}$, with observations. We take molecular cloud data from \citet{HeidermanEtAl2010}, covering the mass range $10$ to $10^4\,M_\odot$ and star formation rates from $\sim\!1$ to $\sim\!30\,M_\odot\,\mathrm{Myr}^{-1}$. From their data we estimate normalised mean star formation rates ranging from $\sim\!10^{-3}$ to $\sim\!10^{-1}$, as indicated by the grey band on panel (c) of figure~\ref{fig:dimless-SFR}. In all cases the average values are consistent with the model in the very early stage of the collapse, i.e., for a very low star formation efficiency.

As a volume-density PDF can not be observed, we need to transform the volume-density PDF into a column-density PDF. This transformation will in general reduce the amount of information, due to projection effects, and so it is not normally possible to restore the volume-density PDF from the column-density PDF, without making assumptions about the underlying geometry of the cloud. This problem has been addressed in detail by, e.g., \citet{BruntEtAl2010}, \citet{BurkhartLazarian2012}, \citet{GinsburgEtAl2013}.

For simplified cases, one can directly compute the volume- and column-density PDFs from the density distribution. Specifically, if we assume a spherically symmetric power-law density distribution with exponent $p$,
\begin{equation}
  \label{eq:spherical-symm-density-profile}
  \rho(r) = \frac{1}{\alpha\,r^{p}},
\end{equation}
there is a logarithmic scaling with the density,
\begin{equation}
  \label{eq:mu}
  \mu\;\equiv\;\frac{\dif \ln V}{\dif\ln\rho} = -\frac{3}{p},
\end{equation}
further details are shown in the appendix. Hence, the slope of the tail of the volume-weighted PDF, $\mu=-1.54$, transforms into $p=-1.95$, which is very close to the density exponent for a singular isothermal sphere, $p=-2$ at the beginning of the collapse (before the rarefaction wave changes the index of the profile to $p=-1.5$ \citep{Shu77}).

For an observer looking along the $z$-axis, the column-density profile is then
\begin{eqnarray}\nonumber
\Sigma(x,y)&=&\int_{-\infty}^{\infty}\rho(x,y,z)\dif z\\\label{eq:column-density-integral}
&=&\frac{\sqrt{\pi}}{\alpha}\,\frac{\Gamma\rkl{\frac{p-1}{2}}}{\Gamma\rkl{\frac{p}{2}}}\,\frac{1}{\rkl{x^2+y^2}^{(p-1)/2}},
\end{eqnarray}
where $\Gamma$ is the Gamma function. In terms of the impact parameter on the observer's sky, $s=\sqrt{x^2+y^2}$, this reduces to
\begin{equation}
  \label{eq:column-density-radial-dependence}
  \Sigma(s) = \frac{1}{\alpha's^{p-1}}.
\end{equation}
From this we can derive the column-density PDF with the corresponding logarithmic scaling of
\begin{align}
  \label{eq:eta}
  \eta\;\equiv\;&\frac{\dif \ln A}{\dif\ln\Sigma} = -\frac{2}{p-1}.
\end{align}

\begin{figure}
  \centering
  \includegraphics[width=8cm]{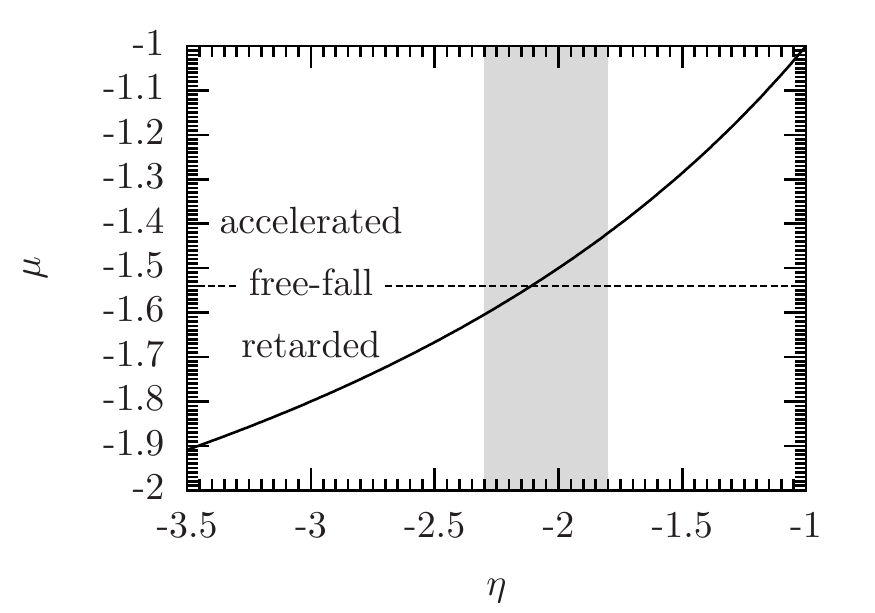}
  \caption{The slope of the volume-density PDF, $\mu$, as a function of the slope of the column-density PDF, $\eta$. The value for pure free-fall, $\mu=-1.54$, corresponds to a slope in the column-density PDF of $\eta=-2.11$. The grey vertical band indicates the range of $\eta$ values found by \citet{KainulainenEtAl2009}.}
  \label{fig:PDF-slopes-conversion-function}
\end{figure}

If we eliminate $p$ between equation~\ref{eq:mu} for $\mu$ (the slope of the high-density tail of volume-density PDF) and equation~\ref{eq:eta} for $\eta$ (the slope of the high-density tail of column-density PDF), we obtain
\begin{equation}
  \mu=\frac{3\eta}{2-\eta};\hspace{1.0cm}\eta\;=\;\frac{2\mu}{3+\mu}.
\end{equation}
Hence, the power-law tail that we predict for the volume-density PDF translates into a power-law tail on the column-density PDF with slope $\eta\!=\!-2.1$. The relationship between $\mu$ and $\eta$ is plotted on figure~\ref{fig:PDF-slopes-conversion-function}. 

\citet{KainulainenEtAl2009} have investigated the structure of molecular clouds, using IR extinction maps, and find a strong correlation between the power-law tail on the column-density PDF and the star formation activity in the cloud. Using their figure~4, we estimate that the tail of the column-density PDF has slope in the range $-2.3\la\eta\la -1.8$. This range is represented by a shaded vertical band on figure~\ref{fig:PDF-slopes-conversion-function}, and corresponds to a volume-density PDF with slope in the range $-1.6\la\mu\la -1.4$, in very good agreement with the prediction of our free-fall analysis.

\citet{SchneiderEtAl2012} have observed the Rosette molecular cloud and extracted column-density PDFs for the entire cloud, and also for individual regions. The global column-density PDF (their figure~6) has a power-law tail with slope $\eta\simeq-3.0$, corresponding to retarded collapse and a volume-density PDF with a power-law tail having slope $\mu\simeq-1.8$. The steep slopes of these column-density and volume-sensity PDFs suggest that the cloud as a whole is not undergoing free-fall collapse, but rather retarded contraction. It is not necessary that the whole cloud be contracting. Splitting the cloud into subregions reveals that the dynamics can be quite differennt from one subregion to another (see their figures~2 and 3). The column-density PDF for the central region (3) shows a flat power-law tail at high column-densities, with slope $\eta\!\simeq\!-1.5$, corresponding to free-fall conditions. In contrast, the column-density PDF for the quiescent region (6) appears not to be contracting at all.

\section{Summary and conclusion}%

We have developed an analytic description for the evolution of the density PDF, based on free-fall collapse of a sphere. We use a simple analytic approximation to describe the collapse in closed functional form. From this we derive the slope of the high-density tail of the density PDF, and the star formation rate as a function of time. Our results can be summarised as follows.
\begin{itemize}
\item In the freefall approximation, the high-density tail of the density PDF asymptotes to a  power-law with slope $\pdf_V(\rho)\propto\rho^{-1.54}$ for the (volume-weighted) density PDF, and $\pdf_M(\rho)\propto\rho^{-0.54}$ for the mass-weighted equivalent. The only condition is that the initial density PDF, at the start of the collapse, have finite slope.
\item The power-law tail appears first at high densities and steadily extends to lower densities as time proceeds. If the initial density PDF at the beginning of the collapse is known, the star formation efficiency can be related to the threshold density at which the high-density tail begins.
\item Physical processes that retard the collapse in marginally unstable density regimes, such as thermal pressure or magnetic fields, steepen the slope of the high-density tail. Although these retarding processes are not explicitly included in our analysis, we can mimic them by varying the parameter $a$ in the approximate collapse solution. Conversely, processes that accelerate the collapse, relative to free-fall, such as driven converging flows, flatten the high-density tail of the PDF.
\item We derive analytic descriptions for the free-fall accretion rate, based on the density PDF. The corresponding star formation rates are highly sensitive to the initial width of the density PDF. Star-forming regions with low turbulent motions, corresponding to narrow density PDFs, lead to retarded star formation and low star formation rates. In contrast, highly turbulent systems, corresponding to broad density PDFs, result in an early and fast star formation process, i.e., a bursting mode of star formation.
\item Comparisons to observations of star-forming regions are in reasonably good agreement with our model. The observed star formation rates are consistent with the model during the early stages of collapse, i.e., for a very low star formation efficiency. Assuming that observed clouds approximate to spherical symmetry, the power-law tails of their column-density PDFs suggests that they are contracting with speeds close to free-fall.
\end{itemize} 

\begin{acknowledgments}
 We thank Steffi Walch and Christoph Federrath for useful comments and inspiring discussions. We also thank the referee for careful reading of the manuscript and detailed comments, which improved the manuscript perceptibly.
 P.G. acknowledges support from the DFG Priority Program 1573 {\em Physics of the Interstellar Medium}.
 P.G. and A.P.W. acknowledge the support of a Marie Curie Research Training Network (MRTN-CT2006-035890), and APW acknowledges the support of an STFC rolling grant (PP/E000967/1).
 P.G. and L.K.~acknowledge support by the International Max Planck Research School for Astronomy and Cosmic Physics (IMPRS) and the Heidelberg Graduate School of Fundamental
 Physics (HGSFP). The HGSFP is funded by the Excellence Initiative of the German Research Foundation DFG GSC 129/1.
 P.G., L.K., and
 R.S.K. acknowledge subsidies from the Baden-W\"{u}rttemberg-Stiftung via contract research in the program {\em Internationale Spitzenforschung II} (grant P-LS-SPII/18).
 R.S.K. furthermore gives thanks for subsidies from the Deutsche Forschungsgemeinschaft (DFG) via the SFB 881 'The Milky Way System' (subprojects B1, B2, and B5), as well as via the SPP 1573 grant (KL 1358/14-1).

\end{acknowledgments}

\begin{appendix}

\section{Thermodynamics}%

\label{sec:thermodynamics}

\begin{figure}
  \centering
  \includegraphics[width=8cm]{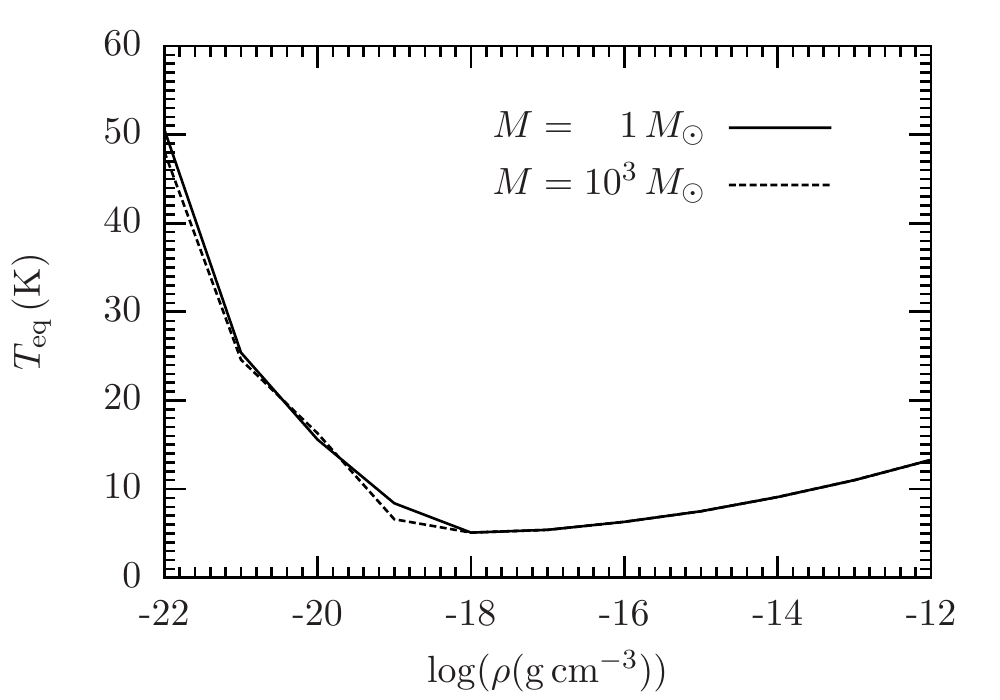}
  \includegraphics[width=8cm]{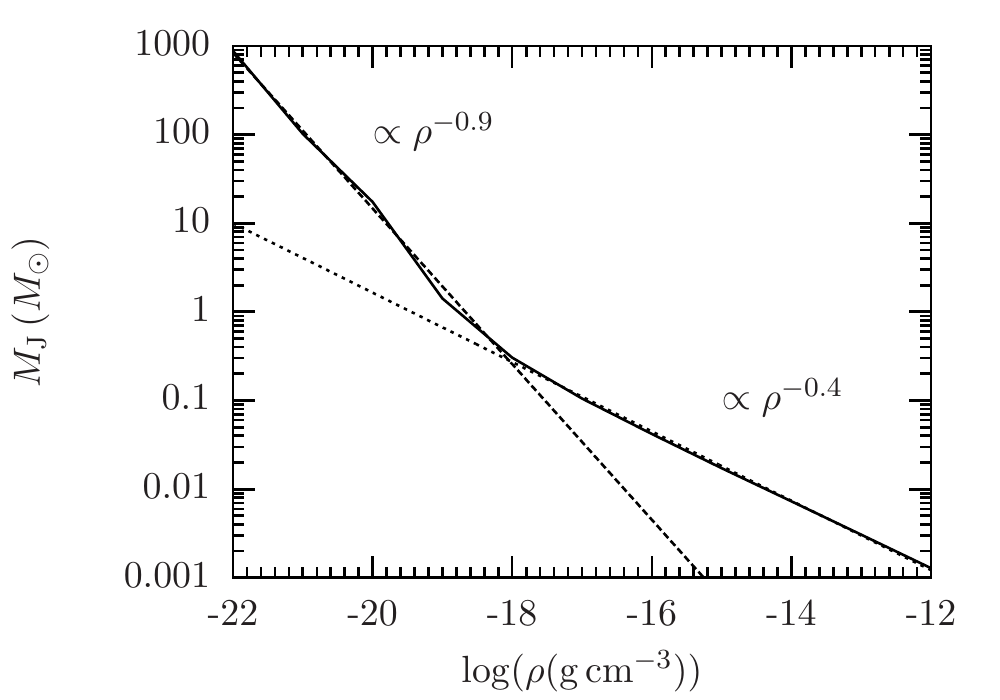}
  \caption{\textsc{Left}: Temperature as a function of density for molecular clouds with solar metallicity \citep[see, e.g.,][]{Kolesnik1973}. \textsc{Right}: resulting scaling of the Jeans mass with density.}
  \label{fig:temperature}
\end{figure}

We can estimate the stability of a cloud with the Jeans analysis. The critical mass above which thermal effects cannot prevent the gravitational collapse is the Jeans mass \citep{Jeans1902},
\begin{equation}
  M_\mathrm{J} = \frac{\pi^2}{6} \rkl{\frac{k_\mathrm{B}}{G\mu m_\mathrm{p}}}^{3/2}\frac{T^{3/2}}{\rho^{1/2}},
\end{equation}
computed as a sphere with the diameter of the Jeans length, $\lambda_\mathrm{J}=\sqrt{\pi k_\mathrm{B} T/ (\mu m_\mathrm{H})}$, where $k_\mathrm{B}$ is Boltzmann's constant, $T$ the temperature, $\mu$ the molecular weight of the gas, $m_\mathrm{H}$ the mass of a hydrogen atom, and $\rho$ the average density of the cloud. The important relation to note is the dependency of the Jeans mass on the temperature and the density. In molecular clouds, the temperature decreases with increasing density as shown in figure~\ref{fig:temperature} up to a density of $\rho\sim10^{-18}\,\gpcc$ and stays roughly constant above that threshold up to a density of $\rho\sim10^{-12}\,\gpcc$. As a result, the Jeans mass decreases with a scaling of $M_\mathrm{J}\propto\rho^{-0.9}$ below $\rho\sim10^{-18}\,\gpcc$ and $M_\mathrm{J}\propto\rho^{-0.4}$ between $10^{-18}\,\gpcc<\rho<10^{-12}\,\gpcc$. For a non-fragmenting cloud with constant mass this means that the cloud becomes more unstable during the collapse and the assumption of free-fall collapse becomes more appropriate during the collapse. If a cloud fragments into many subsystems and the mass of the individual fragments decreases faster due to fragmentation than the density increases due to the collapse, the free-fall approximation eventually breaks. Depending on the degree of fragmentation, our analysis might break down for certain systems. However, the fragmentation of clouds is a subject on its own as it depends on the properties of the turbulence, the physical processes involved, and the initial conditions, which goes beyond the scope of this paper. At this point, we therefore stick to the simplified Jeans analysis and the conclusions concerning the stability.

\section{PDFs for spherically symmetric density distributions}%

For a spherically symmetric power-law density distribution with exponent $p$,
\begin{equation}
  \rho(r) = \frac{1}{\alpha\,r^{p}},
\end{equation}
the (volume-weighted) density PDF takes the form
\begin{equation}
  \label{eq:spherical-symm-PDF-volume}
  \frac{\dif V}{\dif\rho} = -\frac{4\pi}{p}\alpha^{-3/p}\rho^{-3/p-1},
\end{equation}
and mass-weighted equivalent
\begin{equation}
  \label{eq:spherical-symm-PDF-mass}
  \frac{\dif M}{\dif\rho} = -\frac{4\pi}{p}\alpha^{-3/p}\rho^{-3/p}.
\end{equation}
Hence, there is a logarithmic scaling with the density,
\begin{equation}
  \mu\;\equiv\;\frac{\dif \ln V}{\dif\ln\rho} = -\frac{3}{p}, 
\end{equation}
and
\begin{equation}
  \frac{\dif \ln M}{\dif\ln\rho} = -\frac{3}{p}+1,
\end{equation}
respectively. The projected column density distribution reads
\begin{equation}
  \Sigma(s) = \frac{1}{\alpha's^{p-1}}.
\end{equation}
From this we can derive the area-weighted column-density PDF,
\begin{equation}
  \label{eq:column-density-PDF-area-weighted}
  \frac{\dif A}{\dif\Sigma} = -\frac{2\pi}{p-1}\,{\alpha'}^{-2/(p-1)}\,\Sigma^{-2/(p-1)-1},
\end{equation}
and mass-weighted column-density PDF,
\begin{equation}
  \label{eq:column-density-PDF-mass-weighted}
  \frac{\dif M}{\dif\Sigma} = -\frac{2\pi}{p-1}\,{\alpha'}^{-2/(p-1)}\,\Sigma^{-2/(p-1)};
\end{equation}
The corresponding logarithmic scalings are
\begin{equation}
  \eta\;\equiv\;\frac{\dif \log A}{\dif\log\Sigma} = -\frac{2}{p-1},
\end{equation}
and
\begin{equation}
  \frac{\dif \log M}{\dif\log\Sigma} = -\frac{2}{p-1}+1.
\end{equation}
  
\end{appendix}

\bibliographystyle{apj}

\end{document}